\documentstyle[sprocl,epsfig]{article}

\newcommand{\smaf}[2] {{\textstyle \frac{#1}{#2} }}

\begin{document}

\vspace*{-2.5cm}
\begin{flushright}
  UWThPh-1997-46 \\
  HEPHY-PUB 679/97 \\
  hep-ph/9711464 
\end{flushright}
\vspace*{0.6cm}

\title{SUSY--QCD corrections to squark production and decays \\
       in $e^+e^-$ annihilation}

\author{A. Bartl$^{1)}$, H. Eberl$^{2)}$, S. Kraml$^{2)}$, 
        \underline{W. Majerotto}$^{2)\dagger}$, W. Porod$^{1)}$}

\address{$^{1)}$ Institut f\"ur Theoretische Physik der Universit\"at Wien, 
                 Vienna, Austria\\
         $^{2)}$ Institut f\"ur Hochenergiephysik der \"OAW, Vienna, Austria}

\maketitle
\begin{abstract}
We discuss the supersymmetric ${\cal O}(\alpha_s)$ QCD corrections to 
$e^+e^- \to \tilde q_i^{} \bar{\tilde q}_j^{}$ $(i,j = 1,2)$ 
and to $\tilde q_i^{}\to q'\tilde\chi^\pm_j\!$, $q\tilde\chi^0_k$ 
$(i,j=1,2;\,k=1\ldots4)$ within the Minimal Supersymmetric Standard Model.
In particular we consider the squarks of the third generation 
$\tilde t_i$ and $\tilde b_i$ including the left--right mixing.
In the on--shell scheme also the mixing angle has to be renormalized. 
We use dimensional reduction (which preserves supersymmetry) 
and compare it with the conventional dimensional regularization.
A detailed numerical analysis is also presented.
\end{abstract}

\footnotetext{$\dagger$ Talk presented at the 
{\em International Workshop on Quantum Effects in the MSSM}, 
September 9 -- 13, 1997, Barcelona, Spain.}

\setlength{\unitlength}{1mm}   

\section{Introduction}

In supersymmetry (SUSY) one has two types of scalar quarks (squarks),
$\tilde q_L^{}$ and $\tilde q_R^{}$, corresponding to the left and 
right helicity states of a quark. 
$\tilde q_L^{}$ and $\tilde q_R^{}$, however, mix due to the Yukawa coupling
to the Higgs bosons, which is proportional to the mass of the quark. 
One therefore expects large mixing in the case of the stop quarks so that 
one mass eigenstate ($m_{\tilde t_1}$) might be rather light and even 
reachable at present colliders. 
The sbottoms $\tilde b_L,\,\tilde b_R^{}$ may also strongly mix for 
large $\tan\beta$. 
The mass matrix in the basis ($\tilde q_L^{},\,\tilde q_R^{}$) is given by:
\begin{equation}
 {\cal M}^2 =  \left( \begin{array}{lr} 
  m^2_{\tilde Q} + m^2_q + D_1^{} & 
  m_q \left( A_q - \mu \left\{ {\cot\beta \atop \tan\beta} \right\} \right) \\
  m_q \left( A_q - \mu \left\{ {\cot\beta \atop \tan\beta} \right\} \right) &
  m^2_{\tilde U\!,\tilde D} + m^2_q + D_2^{} 
 \end{array} \right) \, .
\label{eq:massmat}
\end{equation}
Here $m_{\tilde Q}$, $m_{\tilde U}$, $m_{\tilde D}$, and $A_q$ are 
SUSY soft--breaking parameters, $\mu$ is the Higgsino mass parameter, 
and $\tan\beta = \frac{v_2}{v_1}$. 
$D_1^{}$ and $D_2^{}$ are the $D$ terms: 
$D_1^{} = m^2_q \cos 2\beta\, (I^{3{\rm L}}_q - e_q \sin^2\theta_W)$, 
$D_2^{} = m^2_Z \cos 2\beta\, e_q \sin^2\theta_W$, 
with $I^{3{\rm L}}_q$ the third component of the weak isospin of $q$.
In the off-diagonal elements of Eq.\,(\ref{eq:massmat}) $\cot\beta$ 
enters in the case of the stops and $\tan\beta$ in that of the sbottoms. 
Diagonalizing the matrix one gets the mass eigenstates
$\tilde q_1^{} = \tilde q_L^{} \cos\theta_{\tilde q} 
                +\tilde q_R^{} \sin\theta_{\tilde q}$, 
$\tilde q_2^{} = -\tilde q_L^{} \sin\theta_{\tilde q} 
                 +\tilde q_R^{} \cos\theta_{\tilde q}$ 
with the masses $m_{\tilde q_1}$, $m_{\tilde q_2}$ 
(with $m_{\tilde q_1} < m_{\tilde q_2}$) 
and the mixing angle $\theta_{\tilde q}$.

Conventional QCD corrections to squark pair production 
$e^+e^- \to \tilde q_i^{} \bar{\tilde q}_j^{}$ $(i,j = 1,2)$ 
can be very large \cite{drees}. 
The SUSY--QCD corrections including squark and gluino exchange will be 
discussed here following closely ref.~\cite{eberl}. 
These corrections were also treated in \cite{arhrib}. 
The new feature in the calculation of the SUSY--QCD corrections 
is that in the on--shell scheme 
a suitable renormalization condition has to be found 
for the mixing angle $\theta_{\tilde q}$ 
because the tree--level amplitude explicitly depends on it. 
We will explain this in detail below. 

The SUSY--QCD corrections to the squark decays into chargino or neutralino,
$\tilde q_i^{} \to q'\tilde\chi^\pm_j$, $q\tilde\chi^0_k$ 
$(i,j=1,2;\:k=1\ldots4)$, have been calculated in \cite{djouadi,kraml} 
and will also be discussed in the following. Here the dependence on the 
nature of the charginos/neutralinos (gaugino--like or higgsino--like) 
is particularly interesting. 

We work in the on--shell scheme and use dimensional reduction 
($\overline{\rm DR}$) 
to regularize the integrals, which is necessary to preserve supersymmetry 
(at least up to two loops). 
We will comment on the differences between this 
and the dimensional regularization scheme used in the Standard Model. 

\section{The Production Process $e^+e^- \to \tilde q_i^{} \bar{\tilde q}_j^{}$}

The cross section at tree level is given by:
\begin{equation}
 \sigma^0 \left( e^+e^- \to \tilde q_i^{} \bar{\tilde{q}}_j \right) =
 \frac{\pi\alpha^2}{s} \lambda^{3/2}_{ij} 
 \left[ e^2_q\, \delta_{ij} 
        - T_{\gamma Z}^{}\, e_q a_{ij} \delta_{ij} 
        + T_{\!Z\!Z}^{}\, a^2_{ij} 
 \right] 
\end{equation}
with \vspace*{-4mm}
\begin{eqnarray}
  T_{\gamma Z}^{} &=& 
    \frac{v_e}{8\, c^2_W s^2_W} \,
    \frac{s(s-m^2_Z)}{\left[(s-m^2_Z)^2 + \Gamma_{\!Z}^2 m_Z^2\right]} \,,\\
  T_{\!Z\!Z}^{} &=& 
    \frac{(a^2_e + v^2_e)}{256\,s^4_W c^4_W} \,
    \frac{s^2}{(s-m^2_Z)^2 + \Gamma_{\!Z}^2 m_Z^2} \, .
\end{eqnarray}

\noindent
Here $\lambda_{ij} = (1 - \mu^2_i - \mu^2_j )^2 - 4 \mu^2_i \mu^2_j$ 
with $\mu^2_{i,j} = m^2_{\tilde q_{i,j}}/ s$. 
$e_q$ is the charge of the squarks (in units of $e$),
$v_e = - 1 + 4 s^2_W$, $a_e = -1$, $s_W = \sin\theta_W$, 
$c_W = \cos\theta_W$. 
$a_{ij}$ are the relevant parts of the couplings 
$Z\tilde q_j^{} \tilde q_i^{\,*}$:
\begin{eqnarray}
  a_{11} &=& 4\,(I^{3{\rm L}}_q \cos^2\theta_{\tilde q} - s^2_W e_q) \,, 
  \quad 
  a_{22}\;=\; 4\,(I^{3{\rm L}}_q \sin^2\theta_{\tilde q} - s^2_W e_q) \,,
  \nonumber \\ 
  a_{12} &=& a_{21} \;=\; -2I^{3{\rm L}}_q \sin 2\theta_{\tilde q} \,.
\end{eqnarray}

The SUSY--QCD corrections in ${\cal O}(\alpha_s)$ consist of the conventional 
QCD corrections \cite{drees} due to gluon exchange and real gluon radiaton, 
as well as of the corrections due to the exchange of a gluino and squarks, 
see Fig.\,1.  
Our input parameters are the physical masses $m_{\tilde q_1}$, 
$m_{\tilde q_2}$, 
$m_{\tilde q}$, $m_{\tilde g}$, and the mixing angle $\theta_{\tilde q}$. 
We use the on--shell sheme where the masses are fixed by the respective
poles of the propagators. 

\clearpage
\begin{figure}[ht]
\centering
\mbox{\epsfig{file=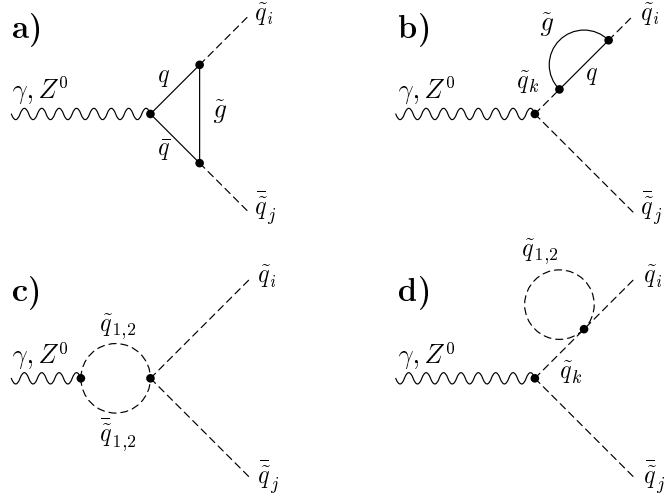,height=7cm}}
\caption{Feynman diagrams for the lowest order SUSY--QCD corrections to
$e^+ e^- \to \tilde q_i^{} \bar{\tilde q}_j^{}$ with squarks and
gluinos in the loop. Note that there are also the corresponding
diagrams to b), and d) for the antisquark $\bar{\tilde q}_j^{}$.}
\end{figure}

\noindent
In renormalizing the lagrangian we follow the usual procedure:
\begin{equation}
  {\cal L}_0 = {\cal L} + \delta{\cal L}
\label{eq:renlag}
\end{equation}
with
\begin{equation}
  {\cal L} = 
   -ee_q\,\delta_{ij}\, A^{\mu}\, \tilde q_i^{\,*} 
      (\stackrel{\leftrightarrow}{i\partial_\mu})\, \tilde q_j^{} 
   -\frac{e}{4 s_W c_W}\,a_{ij}\, Z^{\mu}\, \tilde q_i^{\,*} 
      (\stackrel{\leftrightarrow}{i\partial_\mu})\, \tilde q_j^{}\,.
\end{equation}
${\cal L}_0$, the bare lagrangian, has the same form with 
the bare quantities:
\begin{eqnarray}
  e_q^0\,\delta_{ij} &=& e_q\,\delta_{ij} + (\delta e_q)_{ij} \,, \\
  a_{ij}^0 &=& a_{ij} + \delta a_{ij} \,,\\
  \tilde q_i^{\,* 0}  &=& 
    (1 + \smaf{1}{2} \delta Z_{ii})\, \tilde q_i^{\,*}
    + \delta Z_{ii'}\, \tilde q_{i'}^{\,*} , \quad\; i\neq i' \,, \\
  \tilde q_j^{\,0} &=& 
    (1 + \smaf{1}{2} \delta Z_{jj})\, \tilde q_j^{} 
    + \delta Z_{jj'}\, \tilde q_{j'}^{} , \quad j\neq j' .
  \label{eq:qjbare}
\end{eqnarray}
Notice that because of 
$\theta_{\tilde q}^0 = \theta_{\tilde q} + \delta\theta_{\tilde q}$, 
$\delta a_{ij}$ is a function of $\delta\theta_{\tilde q}$.
The total correction in ${\cal O}(\alpha_s)$ can be written as:
\begin{eqnarray}
  \Delta a_{ij} &=& \delta a_{ij}^{(v )} + \delta a_{ij}^{(w)} 
                  + \delta a_{ij}^{(\tilde\theta)} \,,
     \label{eq:deltaa} \\ 
 (\Delta e_q )_{ij} &=& (\delta e_q )_{ij}^{(v )} 
                      + (\delta e_q )_{ij}^{(w )} \,,
\end{eqnarray}
where $(v)$ denotes the vertex corrections 
and $(w)$ the wave--function corrections. 
The contributions come from gluon, gluino, and squark exchange.
$\delta a_{ij}^{(\tilde\theta)}$ is due to the shift from the bare 
to the on--shell couplings.
As already mentioned, we use dimensional reduction \cite{siegel} 
instead of dimensional regularization. 
Up to first order this is achieved technically by taking 
$D = (4 - r\,\epsilon$) with $r \rightarrow 0$ (see section 4). 
In the case of $e^+e^- \to \tilde q_i^{} \bar{\tilde q}_j^{}$ 
there is, however, no difference between the two schemes 
as will be explained later.

Let us first discuss the vertex corrections $\delta e^{(v)}_{ij}$ 
and $\delta a^{(v)}_{ij}$ coming from the exchange of SUSY particles. 
The gluino contribution due to the graph in Fig.\,1a is given by:
\begin{eqnarray}
  \delta a_{ij}^{(v,\tilde{q})} 
  &=& \frac{2}{3}\frac{\alpha_s}{\pi}\, \Big\{ 
    2m_{\tilde g} m_q v_q S^{\tilde q}_{ij}\, (2C^+_{ij} + C^0_{ij}) 
    \nonumber\\ 
  &+& v_q \delta_{ij} \big[
    (2m_{\tilde g}^2 + 2m_q^2+m_{\tilde q_i}^2 + m_{\tilde q_j}^2)\,C^+_{ij} 
     + 2 m_{\tilde g}^2\, C^0_{ij} + B^0 (s,m^2_q,m^2_q) 
    \big] \nonumber \\
  &+& a_q A^{\tilde q}_{ij} \big[ 
    (2m_{\tilde g}^2 - 2m_q^2 + m_{\tilde q_i}^2 + m_{\tilde q_j}^2)\, C^+_{ij}
    + (m_{\tilde q_i}^2 - m_{\tilde q_j}^2)\, C^-_{ij} 
    \nonumber \\
  & & \hspace{15mm} 
    + 2 m_{\tilde g}^2\,C^0_{ij} + B^0(s,m^2_q,m^2_q) \big] \Big\} 
\end{eqnarray}
and
\begin{eqnarray}
  \delta {(e_q)}_{ij}^{(v, \tilde q)} 
  &=& \frac{2}{3}\frac{\alpha_s}{\pi}\,e_q\, \Big\{ 
    2m_{\tilde g} m_q S^{\tilde q}_{ij}\, (2C^+_{ij}+ C^0_{ij})\,\\
  &+& \delta_{ij}\big[ 
  (2m_{\tilde g}^2 + 2m_q^2 + m_{\tilde q_i}^2 + m_{\tilde q_j}^2) C^+_{ij} 
   + 2m_{\tilde g} C^0_{ij} + B^0(s,m^2_q,m^2_q) 
   \big]\! \Big\} \nonumber
\end{eqnarray}
with $v_q = 2I^{3{\rm L}}_q - 4 s_W^2 e_q$, $a_q = 2I^{3{\rm L}}_q$, 
$S^{\tilde q}_{11}=-\sin 2\theta_{\tilde q}=-S^{\tilde q}_{22} =
 A^{\tilde q}_{12} = A^{\tilde q}_{21}$, and 
$S^{\tilde q}_{12}=S^{\tilde q}_{21}=-\cos 2\theta_{\tilde q} =
 - A^{\tilde q}_{11} = A^{\tilde q}_{22}$.
The functions $C^\pm_{ij}$ are defined by
\begin{equation}
  C^+ = \frac{C^1+C^2}{2}\,, \qquad C^- = \frac{C^1-C^2}{2} \,.
\end{equation}
$B^0$ and $C^{0,1,2}$ are the usual two-- and three--point functions 
as given, for instance, in \cite{denner}.
The arguments of all C--functions are 
$(m_{\tilde q_i}^2, s, m_{\tilde q_j}^2, m_{\tilde g}^2, m_q^2, m_q^2)$. 
%
The squark exchange graph Fig.\,1c is proportional to the 
four--momentum of $Z^0$, and therefore does not contribute to the 
physical matrix element.
The wave--function corrections (Figs.\,1\,b\,,d) can be written as, 
using Eqs.~(\ref{eq:renlag}) to (\ref{eq:qjbare}) $(i\neq i'\!,\, j\neq j')$:
\begin{eqnarray}
  \delta a^{(w)}_{ij} &=& 
    \smaf{1}{2} (\delta Z_{ii} + \delta Z_{jj}) a_{ij} 
    +\delta Z_{i'\!i}\, a_{i'\!j} + \delta Z_{j'\!j}\, a_{ij'} \,.
  \label{eq:dawave}
\end{eqnarray}
An analogous formula holds for $\delta {(e_q)}_{ij}^{(w)}$ with
$a_{ij} \to e_q \,\delta_{ij}$.\\ 
One obtains from Fig.\,1b:
\begin{eqnarray}
  \delta a_{ij}^{(w,\tilde g)} 
  &=& -\mbox{Re} \Big\{ 
    \smaf{1}{2}\big[ \Sigma_{ii}'^{(\tilde g)}(m_{\tilde q_i}^2)
                    +\Sigma_{jj}'^{(\tilde g)}(m_{\tilde q_j}^2) 
               \big]\, a_{ij} \nonumber\\
  & & \hspace{11mm} 
     +\frac{\Sigma_{i'\!i}^{(\tilde g)}(m_{\tilde q_i}^2)}
           {m_{\tilde q_i}^2-m_{\tilde q_{i'}}^2}\, a_{i'j}
     +\frac{\Sigma_{j'\!j}^{(\tilde g)}(m_{\tilde q_j}^2)}
           {m_{\tilde q_j}^2-m_{\tilde q_{j'}}^2}\, a_{ij'}
     \Big\}
  \label{eq:dawsg}
\end{eqnarray}
and
\begin{eqnarray} 
  \delta (e_q)^{(w,\tilde g)}_{ii} 
  &=& -e_q\, \mbox{Re} 
    \left\{ \Sigma_{ii}'^{(\tilde g)}(m_{\tilde q_i}^2) \right\} \, ,  
    \\
  \delta (e_q)^{(w,\tilde g)}_{12} 
  &=& \frac{e_q}{m_{\tilde q_1}^2-m_{\tilde q_{2}}^2}\,\mbox{Re}\left\{ 
    \Sigma_{12}^{(\tilde g)}(m_{\tilde q_2}^2) -
    \Sigma_{21}^{(\tilde g)}(m_{\tilde q_1}^2) \right\} \,,
\end{eqnarray}
where $\Sigma^{(\tilde g)}_{ij} (m^2 )$ are self--energies and 
$\Sigma'^{(\tilde g)}_{ii} (m^2 ) = 
 \partial\Sigma^{(\tilde g )}_{ii}(p^2 ) / \partial p^2 |_{p^{2} = m^{2}}$.
Notice that $\delta (e_q )^{(w,\tilde q)}_{ij} = 0$ because 
the contributions with the squark loop attached at either external 
squark line in Fig.\,1d cancel each other.
The wave--function correction $\delta a_{ij}^{(w,\tilde q)}$ due to Fig.\,1d 
plays an important r$\hat{\rm{o}}$le in the renormalization of the squark 
mixing angle $\theta_{\tilde q}$.

\subsection{Renormalization of the Mixing Angle $\theta_{\tilde q}$}

The total correction $\Delta a_{ij}$, Eq.\,(\ref{eq:deltaa}), 
using Eq.\,(\ref{eq:dawave}) can be written as
$(i\neq i'\!,\, j\neq j')$
\vspace*{-2mm}
\begin{equation}
  \Delta a_{ij} = \delta a^{(v)}_{ij} 
  + \smaf{1}{2} (\delta Z_{ii} + \delta Z_{jj})\, a_{ij} 
  + \delta Z_{i'\!i}\, a_{i'\!j} + \delta Z_{j'\!j}\, a_{ij'} 
  + \delta a^{(\tilde{\theta})}_{ij} \, .
  \label{eq:daij}
\end{equation}
Notice that the first part of the right--hand side, 
$\delta a^{(v)}_{ij} + \frac{1}{2} (\delta Z_{ii} + \delta Z_{jj})\, a_{ij}$, 
is already free of ultra--violet divergencies. 
Hence, the second part of Eq.\,(\ref{eq:daij}) has to be finite, too. 
We therefore may require for $i = 1$ and $j = 2$
\begin{equation}
  \delta a^{(\tilde{\theta})}_{12} =
  (a_{22} - a_{11}) \delta \theta_{\tilde q} = 
  -\left( \delta Z_{21} a_{22} + \delta Z_{12} a_{11}\right) .
  \label{eq:datheta}
\end{equation}
One can easily see that $\Delta a_{ij}$ is then also finite for all $i,\,j$. 
The condition, Eq.\,(\ref{eq:datheta}), means that the {\em non--diagonal} 
self--energy graphs Fig.\,1b and 1d cancel the counterterm 
$\delta a^{(\tilde{\theta})}_{12}$ in Eq.\,(\ref{eq:daij}). 
Notice also that the total squark contribution $\Delta a^{(\tilde q)}_{ij}$ 
is zero. 
Other authors used the same basic idea but took, for instance, the 
condition analogous to Eq.\,(\ref{eq:datheta}) for 
$\delta a^{(\tilde{\theta})}_{11}$ or
$\delta a^{(\tilde{\theta})}_{22}$, see \cite{djouadi}, or 
a similar condition valid at a point $Q^2$, see ref.~\cite{beenakker}. 
The differences between these schemes are, however, numerically very small.

\subsection{Total QCD Correction in ${\cal O} (\alpha_s )$}

The total QCD correction $\Delta\sigma$ to the cross section is
\begin{equation}
  \Delta\sigma = \Delta\sigma^{(g)} + \Delta\sigma^{(\tilde g )} \, ,
\end{equation}
as $\Delta\sigma^{(\tilde q)} = 0$ in our renormalization scheme of the 
squark mixing angle. The gluon contribution factorizes:
\begin{equation}
  \sigma^{(g)} = 
  \sigma^0 \left[ \frac{4}{3}\frac{\alpha_S}{\pi} \Delta_{ij}\right]\,,
\end{equation}
where $\Delta_{ij}$ is given in ref. \cite{eberl}.
The total gluino contribution is given by:
\begin{eqnarray}
  \Delta\sigma^{(\tilde g)} 
  &=& \frac{\pi\alpha^2}{s}\,\lambda^{3/2}_{ij}\, \Big\{ 
    2e_q (\Delta e_q )^{(\tilde g)}_{ij} 
    + 2 T_{\!Z\!Z}^{}\, a_{ij} \Delta a_{ij}^{(\tilde g)} 
    \nonumber \\
  & & \hspace{20mm}
    - T_{\gamma Z}^{} \big[
    e_q \delta_{ij} \Delta a^{(\tilde g)}_{ij} + (\Delta e_q)^{(\tilde g)}_{ij} a_{ij}
    \big] \Big\} 
\end{eqnarray}
with
\begin{eqnarray} 
  \Delta a_{ij}^{(\tilde g)} 
  &=& \delta a_{ij}^{(v,\tilde g)} - \mbox{Re}\,\Big\{ 
    \smaf{1}{2}\big[ \Sigma_{ii}'(m_{\tilde q_i}^2)
                     +\Sigma_{jj}'(m_{\tilde q_j}^2) \big] a_{ij} 
    + \frac{4}{3}\frac{\alpha_s}{\pi}
      \frac{m_{\tilde g} m_q}{m_{\tilde q_1}-m_{\tilde q_2}}\,\delta_{ij}
    \nonumber \\   
  & & \cdot\,\Big[ B^0(m_{\tilde q_i}^2,m_{\tilde g}^2,m_q^2) \,
    \big[ (-1)^{i+1}\, 2 a_{ii'} \cos 2\theta_{\tilde q} 
           - a_{i'i'} \sin 2\theta_{\tilde q} \big] 
    \nonumber \\   
  & & \quad\; +\,B^0(m_{\tilde q_i}^2,m_{\tilde g}^2,m_q^2)\, 
              a_{ii} \sin 2\theta_{\tilde q} 
    \Big] \Big\}  
\end{eqnarray}
$(i\neq i')$ and 
$\Delta (e_q )_{ij}^{(\tilde g)} = 
 (\delta e_q )^{(v)}_{ij} + (\delta e_q)_{ij}^{(w)}$.

\subsection{Discussion} 

First, we have calculated the SUSY--QCD corrections to the cross section of 
$e^+e^- \to \tilde t_1 \bar{\tilde t}_1$ in the LEP energy range 
$\sqrt{s}\leq 200$~GeV. 
We have found that, whereas the conventional QCD correction may be rather 
large, the gluino correction is only about 1\% of the tree--level cross 
section, quite independent of $m_{\tilde{t}_1}$. 
The correction due to gluino exchange is, however, not negligible ($2-8\%$) 
in the energy range of a linear $e^+e^-$ collider 
($\sqrt{s} = 500 - 2000$~GeV).

The $\sqrt{s}$--dependence of the SUSY--QCD corrections to the cross section 
$\sigma (e^+e^- \to \tilde t_1 \bar{\tilde t}_1)$ is shown in Fig.\,2 
for $m_{\tilde t_1}=100$~GeV, $m_{\tilde t_2}= 400$~GeV, 
$m_{\tilde g}=300$~GeV, and $\cos\theta_{\tilde{t}} = 1/\sqrt{2}$. 
The peak at $\sqrt{s}=350$~GeV is due to the $t\bar t$ threshold. 
In Fig.\,3 we show the $\cos\theta_{\tilde{t}}$ dependence of the 
corrections for this process at $\sqrt{s} = 500$~GeV for the same 
masses of the stops and the gluino as in Fig.\,2. 
Whereas the gluon correction has the same behaviour in $\theta_{\tilde{t}}$ 
as the tree--level cross section, the gluino correction is different.
In Figs.~4 and 5 we exhibit the corrections to
$\sigma (e^+e^- \to \tilde t_1 \bar{\tilde t}_2$) and 
$\sigma (e^+e^- \to \tilde t_2 \bar{\tilde t}_2$), respectively, 
at $\sqrt{s}=2$~TeV for $m_{\tilde t_1}=400$~GeV, $m_{\tilde t_2}=800$~GeV,
$m_{\tilde g}=600$~GeV. The gluino contributions can go up to about $-10$\%. 
Fig.\,6 shows the dependence on the gluino mass. 
It is interesting to notice that the gluino correction decreases 
very slowly with the gluino mass. 

\clearpage

\noindent
\refstepcounter{figure}
\begin{minipage}[t]{57mm} 
\begin{picture}(57,50)
\put(0,-7){\mbox{\epsfig{file=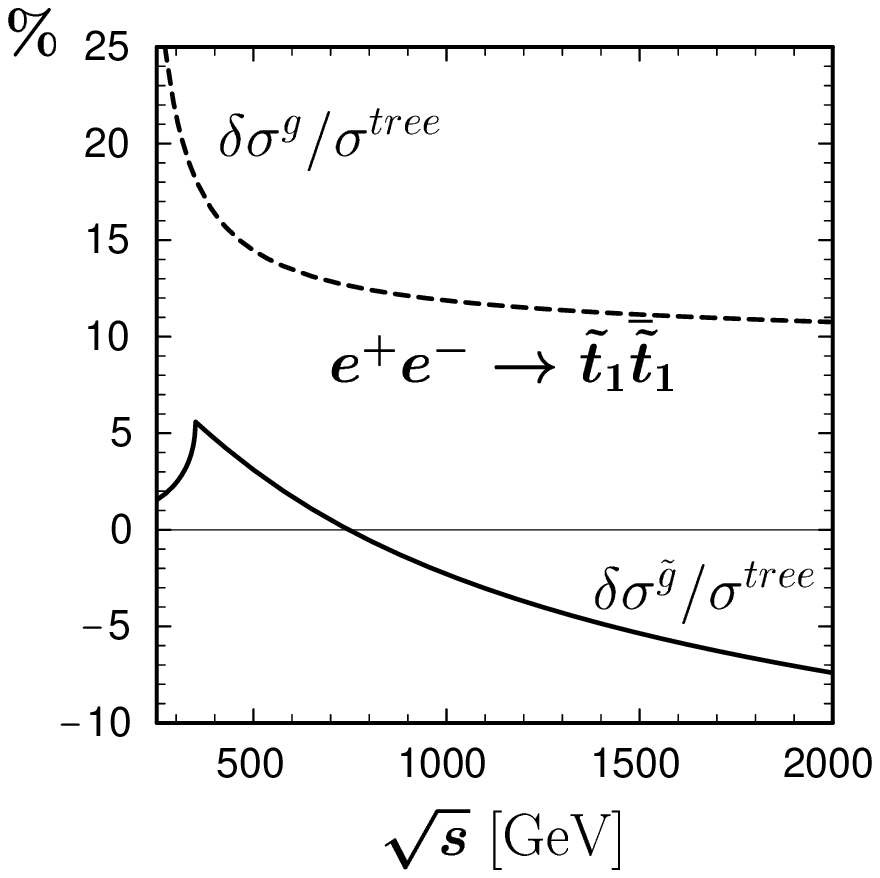,height=5.5cm}}}
\end{picture} 
\vspace{4mm} \\
{\footnotesize \mbox{Figure~\arabic{figure}:}
SUSY--QCD corrections \\ $\delta\sigma^g/\sigma^{tree}$ and
$\delta\sigma^{\tilde g}/\sigma^{tree}$
for $e^+ e^- \to \tilde t_1 \bar{\tilde t}_1$
as a function of $\sqrt{s}$ for $\cos\theta_{\tilde t} = 1/\sqrt{2}$,
$m_{\tilde t_1} = 100$~GeV, $m_{\tilde t_2} = 400$~GeV, and
$m_{\tilde g} = 300$~GeV.}
\end{minipage}
\hspace{4mm}
\refstepcounter{figure}
\begin{minipage}[t]{57mm} 
\begin{picture}(57,50)
\put(0,-6){\mbox{\epsfig{file=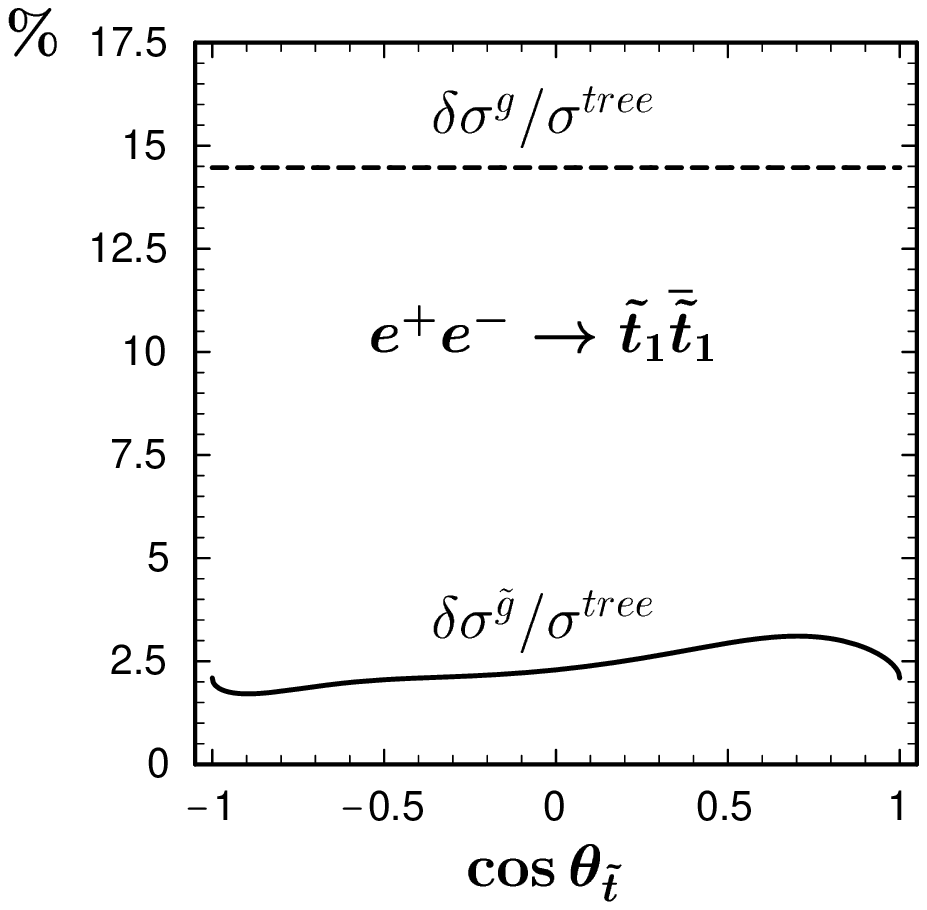,height=5.2cm}}}
\end{picture} 
\vspace{4mm} \\
{\footnotesize \mbox{Figure~\arabic{figure}:} 
SUSY--QCD corrections \\ $\delta\sigma^g/\sigma^{tree}$ and
$\delta\sigma^{\tilde g}/\sigma^{tree}$
for $e^+ e^- \to \tilde t_1 \bar{\tilde t}_1$
as a function of $\cos\theta_{\tilde t}$ for
$\sqrt{s} = 500$~GeV, $m_{\tilde t_1} = 100$~GeV, 
$m_{\tilde t_2} = 400$~GeV, and $m_{\tilde g} = 300$~GeV.}
\end{minipage}
\vspace{5mm} \\
\noindent
\refstepcounter{figure}
\begin{minipage}[t]{57mm}
\begin{picture}(57,50)
\put(0,-7){\mbox{\epsfig{file=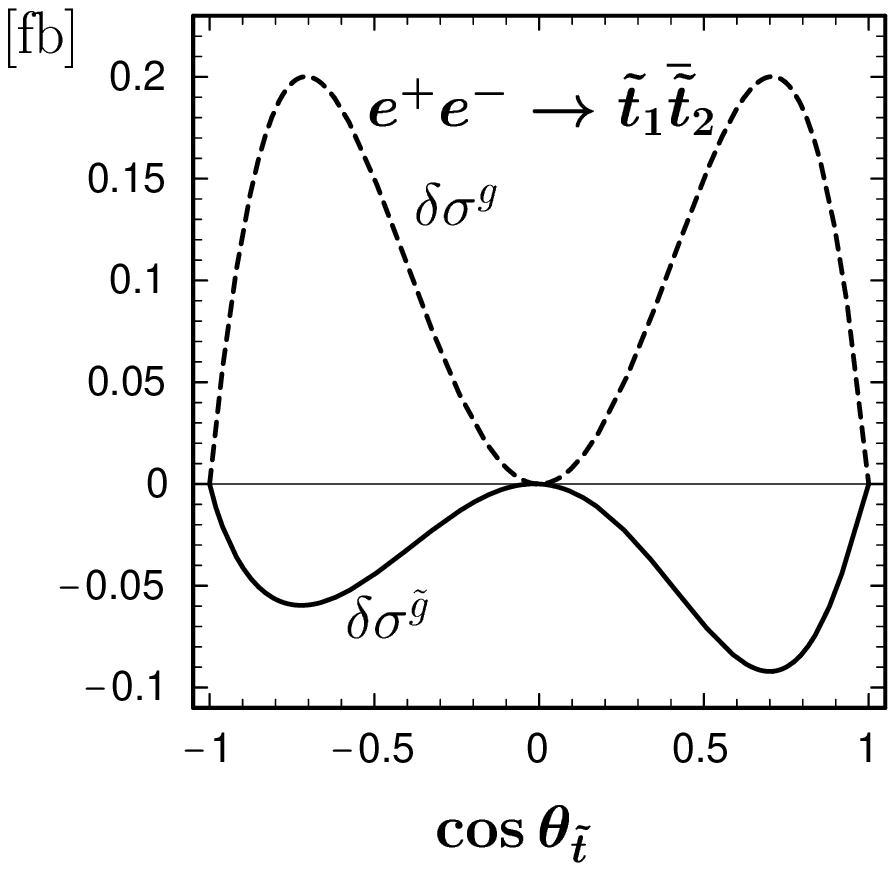,height=5.5cm}}}
\end{picture}
\vspace{4mm} \\
{\footnotesize \mbox{Figure~\arabic{figure}:}
SUSY--QCD corrections $\delta\sigma^g$ and $\delta\sigma^{\tilde g}$ 
as a function of $\cos\theta_{\tilde t}$ 
for $e^+ e^- \to \tilde t_1 \bar{\tilde t}_2$,
$\sqrt{s}=2$~TeV, $m_{\tilde t_1}=400$~GeV, $m_{\tilde t_2}=800$~GeV, 
and $m_{\tilde g}=600$~GeV.}
\end{minipage}
\hspace{4mm}
\refstepcounter{figure}
\begin{minipage}[t]{57mm}
\begin{picture}(57,50)
\put(0,-6){\mbox{\epsfig{file=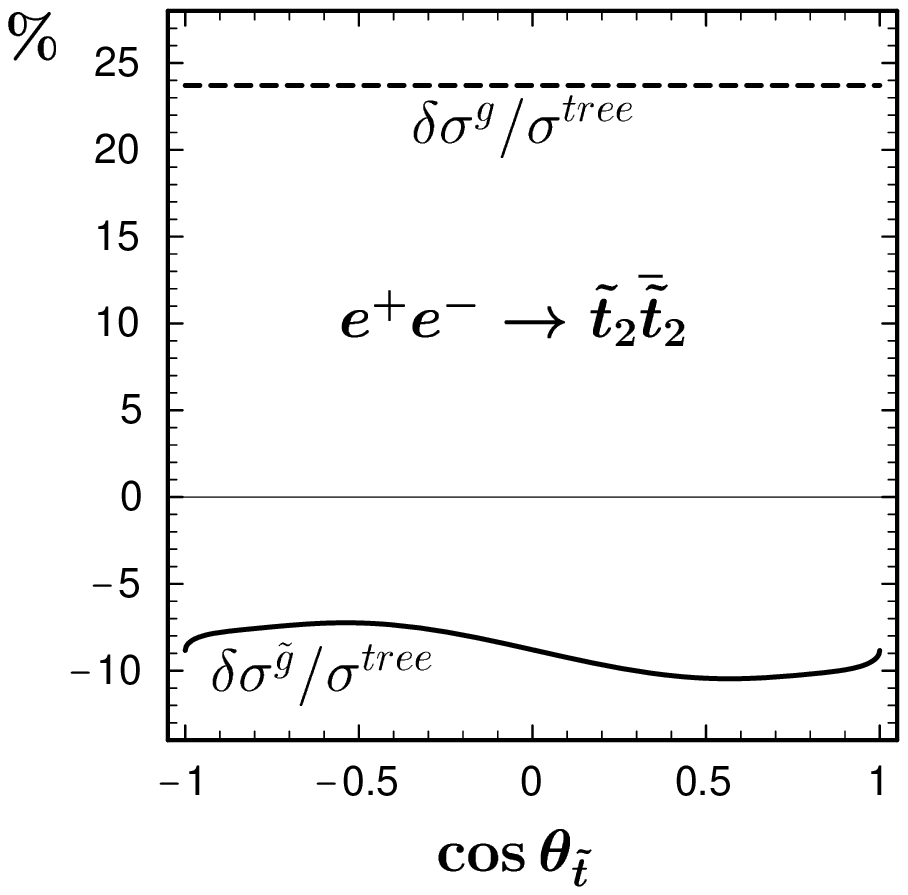,height=5.2cm}}}
\end{picture}
\vspace{4mm} \\
{\footnotesize \mbox{Figure~\arabic{figure}:}
SUSY--QCD corrections \\ $\delta\sigma^g/\sigma^{tree}$ and
$\delta\sigma^{\tilde g}/\sigma^{tree}$
for $e^+ e^- \to \tilde t_2 \bar{\tilde t}_2$
as a function of $\cos\theta_{\tilde t}$ for
$\sqrt{s}=2$~TeV, $m_{\tilde t_1}=400$~GeV, $m_{\tilde t_2}=800$~GeV, 
and $m_{\tilde g} = 600$~GeV.}
\end{minipage}

\clearpage

\noindent
\refstepcounter{figure}
\begin{minipage}[b]{65mm} 
\begin{picture}(65,53)
\put(0,0){\mbox{\epsfig{file=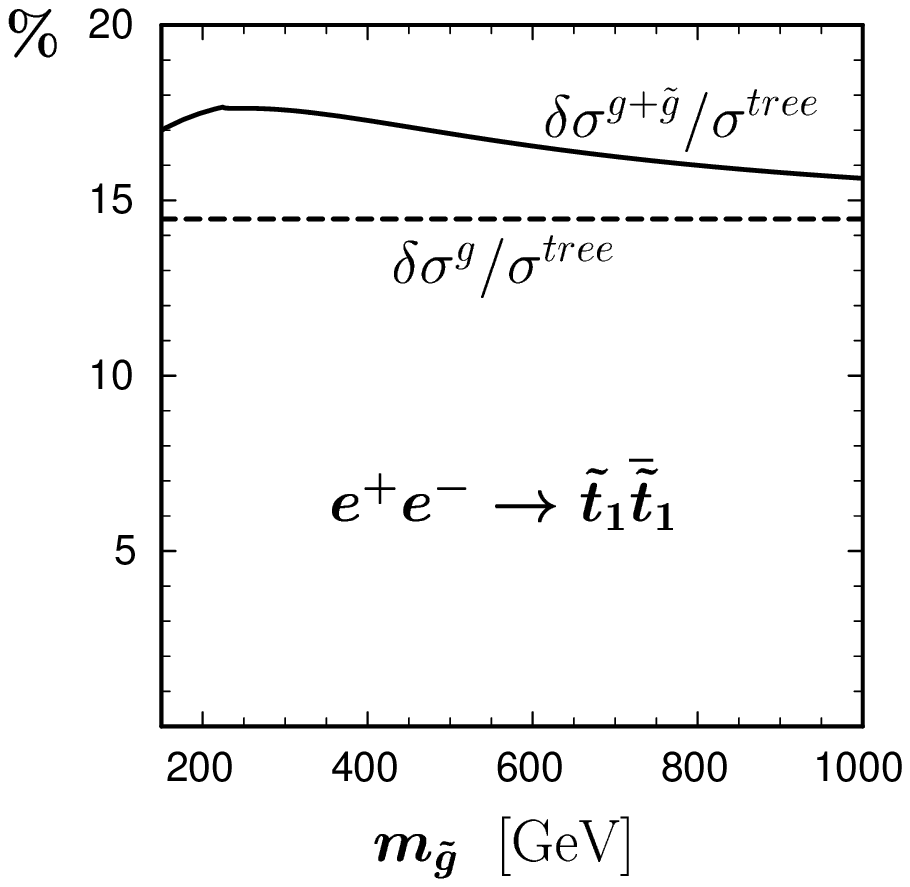,height=5.5cm}}}
\end{picture} 
\end{minipage} 
\begin{minipage}[b]{55mm} 
{\footnotesize \mbox{Figure~\arabic{figure}:}
Dependence of the SUSY--QCD corrections $\delta\sigma^g/\sigma^{tree}$ and
$\delta\sigma^{g+\tilde g}/\sigma^{tree}$
on the gluino mass for $e^+ e^- \to \tilde t_1 \bar{\tilde t}_1$,
for $\sqrt{s} = 500$~GeV, $m_{\tilde t_1} = 100$~GeV, $m_{\tilde t_2} = 400$~GeV,
$\cos\theta_{\tilde t} = 1/\sqrt{2}$.}\\[14mm]
\end{minipage}

\section{Squark Decays into Charginos and Neutralinos}

In the following we discuss the SUSY--QCD corrections for the decays:
\begin{eqnarray}
  \tilde t_i \;\to\; b\,\tilde\chi^+_j , & &
  \tilde b_i \;\to\; t\,\tilde\chi^-_j ,
  \\
  \tilde t_i \;\to\; t\,\tilde\chi^0_k \,, & & 
  \tilde b_i \;\to\; b\,\tilde\chi^0_k \,, 
\end{eqnarray}
with $i,j = 1,2$ and $k = 1\ldots 4$.
The supersymmetric QCD corrections were calculated for $m_q = 0$ and
$\tilde\chi^0_1$ being a photino in ref.~\cite{hikasa}, 
and taking into account squark mixing, quark masses (i.e. Yukawa couplings), 
and general gaugino--higgsino mixing of charginos and neutralinos 
in refs.~\cite{djouadi}~and~\cite{kraml}. 
The decay width at tree--level for $\tilde t_i \to b \tilde\chi^+_j$ 
is given by:
\begin{eqnarray}
  \Gamma^0 (\tilde{t}_i \to b \tilde\chi^+_j ) 
  &=& \frac{g^2 \kappa (m^2_{\tilde t_i},m^2_b,m^2_{\tilde\chi^+_j})}
           {16\pi m^3_{\tilde t_i}} \nonumber \\
  & & \cdot \left( 
  \big[ (\ell^{\,\tilde t}_{ij})^2 + (k^{\tilde t}_{ij})^2 \big]\, X
  - 4\,\ell^{\,\tilde t}_{ij} k^{\tilde t}_{ij} m_b m_{\tilde\chi^+_j} \right) 
\end{eqnarray}
with $X = m^2_{\tilde t_i} - m^2_b - m^2_{\tilde\chi^+_j}$ and
$\kappa(x,y,z) = [(x-y-z)^2 - 4 y z]^{1/2}$. 
The $\tilde{t}_i^*$-$b$-$\tilde\chi^+_j$ couplings 
$\ell^{\,\tilde t}_{ij}$ and $k^{\tilde t}_{ij}$ read, 
for instance, for $\tilde{t}_1 \to b \tilde\chi^+_j$:
\begin{eqnarray}
  \ell^{\,\tilde t}_{1j} &=& 
    - V_{j1} \cos\theta_{\tilde t} 
    + \frac{m_t}{\sqrt{2}\,m_W\sin\beta}\, V_{j2}\sin\theta_{\tilde t}\,,\\
  k^{\tilde t}_{1j} &=& 
  \frac{m_b}{\sqrt{2}\,m_W\cos\beta}\, U_{j2}\cos\theta_{\tilde{t}} \,,
\end{eqnarray}
where $U$ and $V$ are the matrices diagonalizing the charged
gaugino--higgsino mass matrix \cite{haber}.

The ${\cal O}(\alpha_s)$ SUSY--QCD corrected decay width can be written as:
\begin{equation}
  \Gamma = \Gamma^0 + \delta\Gamma^{(v)} + \delta\Gamma^{(w)} 
  + \delta\Gamma^{(c)} + \delta\Gamma^{({\rm real\,gluon})} ,
\label{eq:gammacorr}
\end{equation}
where the superscript $v$ again denotes the vertex correction (Figs.\,7\,a,b) 
and $w$ the wave--function correction (Figs.\,7\,c-g). 
$\delta\Gamma^{(c)}$ corresponds to the shift from the bare 
to the on--shell couplings, taking into account the renormalization 
of the quark mass and the squark mixing angle. 
$\delta\Gamma^{({\rm real\,gluon})}$ is the correction due to real gluon 
bremsstrahlung and cancels the infrared divergencies. \\[1mm]

\noindent
\refstepcounter{figure} 
\begin{picture}(100,75)
\put(5,0){\mbox{\epsfig{file=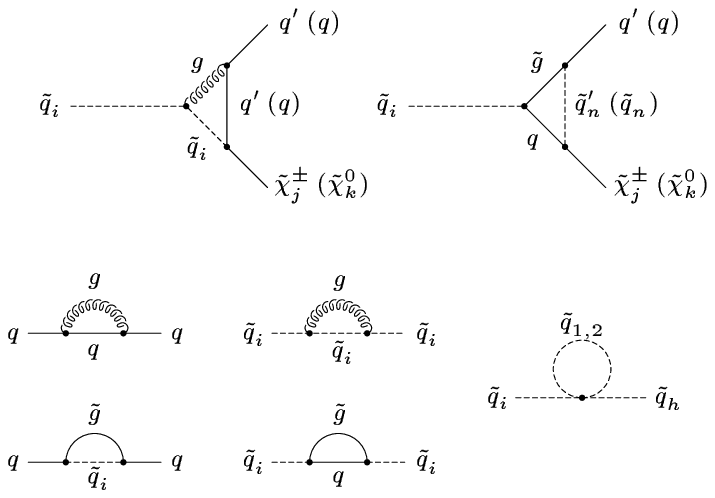,width=110mm}}}
\put(11,72){\makebox(0,0)[bl]{\bf a)}}
\put(65,72){\makebox(0,0)[bl]{\bf b)}}
\put(6,34){\makebox(0,0)[bl]{\bf c)}}
\put(6,14){\makebox(0,0)[bl]{\bf d)}}
\put(43,34){\makebox(0,0)[bl]{\bf e)}}
\put(43,14){\makebox(0,0)[bl]{\bf f)}}
\put(80,26){\makebox(0,0)[bl]{\bf g)}}
\end{picture} \\[2mm]
\noindent
{\footnotesize \mbox{Figure~\arabic{figure}:}
Vertex and wave-function corrections to squark de\-cays into charginos 
and neutralinos. } \\

\noindent
The procedure of the calculation is completely analogous to that discussed 
just before in section 2. 
The complete formulae for the different correction parts in 
Eq.\,(\ref{eq:gammacorr}) are given in ref. \cite{kraml}. 
We want to note that, contrary to the production process 
$e^+e^- \to \tilde q_i^{}\bar{\tilde q}_j$, the corrections to the decay 
widths of $\tilde q_i \to q' \tilde\chi^\pm_j$ and 
$\tilde q_i^{} \to q \tilde\chi^0_k$ are different in the 
dimensional regularization and in the dimensional reduction scheme.
(At first order the difference is finite.) 
This is because of the quark wave--function correction due to 
gluon exchange, Fig.\,7c.  
The quark self--energy corresponding to Fig.\,7c is given by:
\begin{equation}
  \Pi^{(g)} (k^2 ) = \frac{\alpha_s}{3\pi} 
  \left[\, 2/\hspace{-1.8mm}k B^1 + 2 (/\hspace{-1.8mm}k - 2m_q) B^0 
          -r(/\hspace{-1.8mm}k - 2m_q) \right]
\label{eq:pik}
\end{equation}
with $B^n = B^n (k^2,\lambda^2,m^2_q)$ and the gluon mass $\lambda\to 0$.
This leads to the quark wave--function renormalization constants due 
to gluon exchange
\begin{equation}
  \delta Z^{L(g)} = \delta Z^{R(g)} = 
  -\frac{2}{3} \frac{\alpha_s}{\pi} 
  \left[ B^0 + B^1 - 2m^2_q ({\dot B}^0 - {\dot B}^1) -\frac{r}{2}\,\right] 
\label{eq:zet}
\end{equation} 
with $B^n = B^n (m^2_q,\lambda^2,m^2_q)$, 
${\dot B}^n = {\dot B}^n (m^2_q,\lambda^2,m^2_q)$. 
$\delta Z^L$ and $\delta Z^R$ are defined by the usual relation between 
the unrenormalized quark field $q^0$ and the renormalized one, 
$q^0 = (1 + \frac{1}{2}\delta Z^L P_{\!L}^{} 
          + \frac{1}{2}\delta Z^R P_{\!R}^{})\,q$. 
Note the dependence on $r$ in Eqs.~(\ref{eq:pik}) and (\ref{eq:zet}), 
where $r=0$ in the dimensional reduction and $r=1$ in the dimensional 
regularization scheme. 
Note, however, that there is no such difference for the squark 
self--energy graph due to gluon exchange. 

\subsection{Numerical Results}

Let us first discuss the decay $\tilde t_1 \to b\tilde\chi^+_1$, where we 
take $m_{\tilde\chi_1^+}=100$~GeV, $\tan\beta=2$, $m_{\tilde t_2}=600$~GeV, 
$m_{\tilde b_1}=450$~GeV, $m_{\tilde b_2}=470$~GeV, and 
$\cos\theta_{\tilde b}=-0.9$.  We study three cases: 
$M \ll  |\mu|$~($M= 95$~GeV, $\mu=-800$~GeV), 
$M \sim |\mu|$~($M=100$~GeV, $\mu=-100$~GeV), and 
$M \gg  |\mu|$~($M=300$~GeV, $\mu= -89$~GeV). 
We use the GUT relations: $M' \simeq 0.5$~M, $m_{\tilde g} \simeq 3.5$~M.

In Fig.\,8 the dependence of the SUSY--QCD corrections on the stop mass 
is exhibited for $\cos\theta_{\tilde{t}} = 0.6$. 
Notice the pronounced dependence on the nature of the chargino. 
The corrections are largest ($\sim -25\%$), if the chargino is
higgsino--like ($|\mu| \ll M$) due to the large top Yukawa coupling. 
If $\tilde\chi^+_1$ is gaugino--like ($M \ll |\mu|$) the corrections are 
between $+20\%$ and $-10\%$. 

In Fig.\,9 we show the SUSY--QCD corrected widths together with the 
tree--level widths as a function of $\cos\theta_{\tilde{t}}$ for 
$m_{\tilde t_1}=200$~GeV and the other parameters as in Fig.\,8. 
Again, the corrections are biggest in the case of a higgsino--like chargino. 
The behaviour of the $\cos\theta_{\tilde t}$ dependence reflects the fact that 
if $\tilde t_1 \sim \tilde t_R$~$(\cos\theta_{\tilde t}\sim 0)$ it strongly 
couples to the higgsino component of $\tilde\chi^+_1$, 
and if $\tilde t_1 \sim \tilde t_L$~($\cos\theta_{\tilde t}\sim\pm1$) 
it strongly couples to the gaugino component.

In Fig.\,10 we show $\delta\Gamma /\Gamma^0$ [\%] as a function of 
$m_{\tilde t_1}$ for $\tilde t_1 \to t\tilde\chi^0_1$, taking 
$m_{\tilde\chi^0_1}=80$~GeV, $\tan\beta = 2$, $m_{\tilde t_2}=600$~GeV,  
and $\cos\theta_{\tilde t}=0.6$. 
Again we observe that if $\tilde\chi^0_1$ is higgsino--like 
($|\mu| \ll M$) the corrections are about $-20\%$.

We have also studied the dependence on the gluino mass. 
In Fig.\,11 we show a plot where $\delta\Gamma /\Gamma^0$ is exhibited 
for $\tilde t_1 \to b\tilde\chi^+_1$ and $\tilde t_1 \to t\chi^0_1$ 
as a function of $m_{\tilde g}$ for $m_{\tilde t_1}=300$~GeV, 
$\cos\theta_{\tilde t}=0.6$, $\tan\beta = 2$, and $\mu=-100$ and $-800$~GeV. 
$M$ is fixed by $M\simeq 0.3\,m_{\tilde g}$. 
Notice that the SUSY--QCD corrections are still important for 
$m_{\tilde g}\sim 1$~TeV and no decoupling of the gluino mass can be seen. 
This is also the case if we relax the condition $M \simeq 0.3\,m_{\tilde g}$ 
and keep the chargino (neutralino) mass fixed.

\clearpage

\noindent
\refstepcounter{figure}
\begin{minipage}[t]{57mm} 
\begin{picture}(50,54)
\put(2,0){\mbox{\epsfig{file=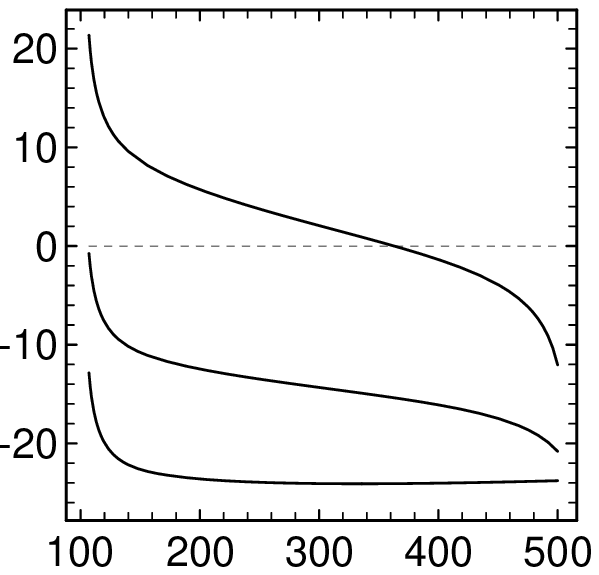,width=5.1cm,height=5.1cm}}}
\put(2,51){\makebox(0,0)[bl]{{\boldmath $\delta\Gamma/\Gamma^0$}~[\%]}}
\put(31,0){\makebox(0,0)[tc]{{\boldmath $m_{\tilde t_1}$}~[GeV]}}
\put(21.5,34.5){\makebox(0,0)[bl]{{\small $(95,-800)$}}}
\put(28,18){\makebox(0,0)[bl]{{\small $(100,-100)$}}}
\put(16,11){\makebox(0,0)[bl]{{\small $(300,-89)$}}}
\end{picture} 
\vspace{4mm} \\
{\footnotesize \mbox{Figure~\arabic{figure}:}  
  SUSY--QCD corrections to the width of 
  $\tilde t_1^{} \to b \tilde\chi_1^+$ 
  as a function of $m_{\tilde t_1}$,
  for $m_{\tilde \chi_1^+}=100$ GeV, $\cos\theta_{\tilde t}=0.6$, 
  $\tan\beta=2$, and various $(M,\mu)$ [GeV] values.}
\end{minipage}
\hspace{4mm}
\refstepcounter{figure}
\begin{minipage}[t]{57mm} 
\begin{picture}(50,54)
\put(3.5,1.8){\mbox{\epsfig{file=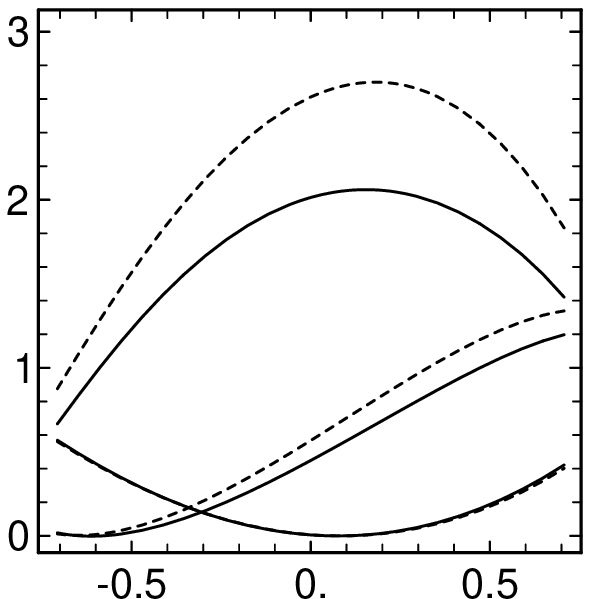,width=4.8cm,height=4.8cm}}}
\put(2.8,51.3){\makebox(0,0)[bl]{{\boldmath $\Gamma$}~[GeV]}}
\put(30,0){\makebox(0,0)[tc]{\boldmath $\cos\theta_{\tilde t_1}$}}
\put(28,9){\makebox(0,0)[bl]{{\small $(95,-800)$}}}
\put(39,22){\makebox(0,0)[br]{{\small $(100,-100)$}}}
\put(34,37){\makebox(0,0)[bc]{{\small $(300,-89)$}}}
\end{picture} 
\vspace{4mm} \\
{\footnotesize \mbox{Figure~\arabic{figure}:}
  Tree--level (dashed lines) and SUSY--QCD corrected (solid lines)
  decay widths of $\tilde t_1^{} \to b \tilde\chi_1^+$ 
  as a function of $\cos\theta_{\tilde t}$, 
  for $m_{\tilde t_1}=200$ GeV, $m_{\tilde \chi_1^+}=100$ GeV, 
  $\tan\beta=2$, and various $(M,\mu)$ [GeV] values.}
\end{minipage} 
\vspace{12mm}\\ 
\noindent
\refstepcounter{figure}
\begin{minipage}[t]{57mm} 
\begin{picture}(50,50)
\put(2,0){\mbox{\epsfig{file=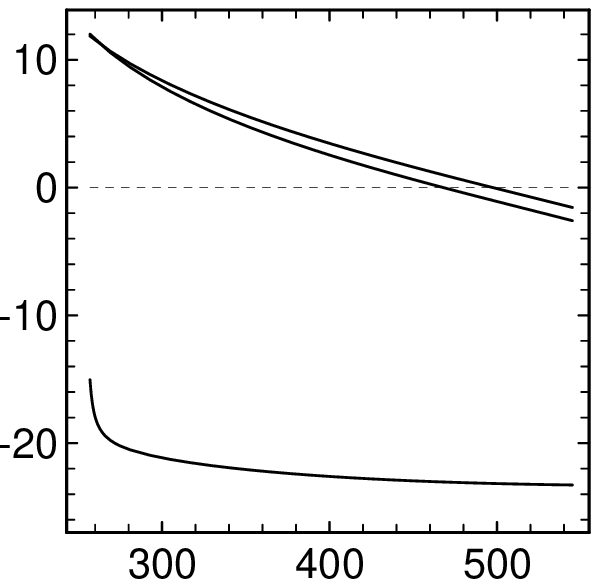,width=5cm,height=5cm}}}
\put(2,51){\makebox(0,0)[bl]{{\boldmath $\delta\Gamma/\Gamma^0$}~[\%]}}
\put(31,0){\makebox(0,0)[tc]{{\boldmath $m_{\tilde t_1}$}~[GeV]}}
\put(26,40){\makebox(0,0)[bl]{{\small $(158,-800)$}}}
\put(47,31){\makebox(0,0)[tr]{{\small $(154,-150)$}}}
\put(18,12){\makebox(0,0)[bl]{{\small $(300,-85)$}}}
\end{picture} 
\vspace{4mm} \\
{\footnotesize \mbox{Figure~\arabic{figure}:} 
  SUSY--QCD corrections to the width 
  of $\tilde t_1^{} \to t \tilde\chi_1^0$ 
  as a function of $m_{\tilde t_1}$,
  for $m_{\tilde \chi_1^0}=80$~GeV, $\cos\theta_{\tilde t}=0.6$, 
  $\tan\beta=2$, and various $(M,\mu)$ [GeV] values.}
\end{minipage}
\hspace{4mm}
\refstepcounter{figure}
\begin{minipage}[t]{57mm} 
\begin{picture}(50,50)
\put(2,-1.2){\mbox{\epsfig{file=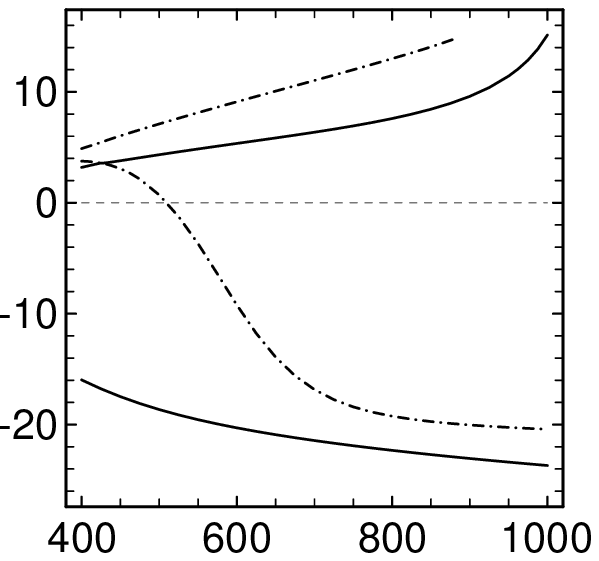,width=5.24cm,height=5.24cm}}}
\put(2,51){\makebox(0,0)[bl]{{\boldmath $\delta\Gamma/\Gamma^0$}~[\%]}}
\put(31,0){\makebox(0,0)[tc]{{\boldmath $m_{\tilde g}$}~[GeV]}}
\put(28.5,33){\makebox(0,0)[bl]{{\small $\mu=-800\,$GeV}}}
\put(28.5,22){\makebox(0,0)[bl]{{\small $\mu=-100\,$GeV}}}
\put(38,36){\vector(0,1){3}}
\put(37,36){\vector(-1,1){6.2}}
\put(39,21){\vector(0,-1){7.3}}
\put(38,21){\vector(-1,-2){5}}
\end{picture} 
\vspace{4mm} \\
{\footnotesize \mbox{Figure~\arabic{figure}:}
  SUSY--QCD corrections to the widths 
  of $\tilde t_1^{} \to b \tilde\chi_1^+$ (solid lines) 
  and $\tilde t_1^{} \to t \tilde\chi_1^0$ (dash-dotted lines)
  as a function of $m_{\tilde g}$,
  for $m_{\tilde t_1}=300$~GeV, $\cos\theta_{\tilde t}=0.6$, 
  $\tan\beta=2$, $M\sim 0.3\,m_{\tilde g}$.}
\end{minipage}

\section{Dimensional Reduction Technique}

The regularization by dimensional reduction was proposed by \cite{siegel}. 
It means that only the space--time dimensions 
(the coordinates $x^\mu$ and momenta $p^\mu$) are continued to 
$D = 4-\epsilon$ dimensions, whereas the vector fields and spinors remain 
four--dimensional.
Following \cite{capper} it is convenient to write the four--dimensional 
vector field $V_\mu$ as $V_\mu = (V_i,\,V_\sigma)$, where $V_i$ is a 
$D$--dimensional vector, and $V_\sigma$ is $\epsilon$--dimensional behaving 
as $\epsilon$ scalars. 
Moreover, one has $\gamma^\mu = (\gamma^i,\gamma^\sigma)$.
Note that $x^\mu = (x^i, 0)$, $\partial^\mu = (\partial^i, 0)$, and 
$p^\mu = (p^i, 0)$. 
As a consequence the lagrangian ${\cal L}$ can be decomposed as 
${\cal L} = {\cal L}^{(D)} + {\cal L}^{(\epsilon)}$, where ${\cal L}^{(D)}$ 
is the lagrangian of the conventional dimensional regularization.
Therefore, to each interaction term of a vector field there is a corresponding
``$\epsilon$ scalar'' interaction term, 
except for the vector--scalar--scalar interaction because 
\mbox{$V^\mu\,\phi^{*}\!\stackrel{\leftrightarrow}{\partial}_\mu\!\phi = 
 V^i\,\phi^{*}\!\stackrel{\leftrightarrow}{\partial}_i\!\phi$} 
(and no $\epsilon$ term), with $\phi$ being a scalar field. 
Therefore, in this case there is no difference between dimensional
regularization and dimensional reduction. 
There is, however, a difference in the case of the interaction of a fermion 
with a vector field. For instance, the fermion self--energy, Fig.\,7c, 
receives a contribution due to $\epsilon$ scalars in the loop of 
$\frac{\alpha_s}{3\pi}(/\hspace{-1.8mm}k-2m_q)$. 
This is just the expression which cancels the $r$--dependent term 
in Eq.\,(\ref{eq:pik}) for $r=1$ in order to get the result of  
dimensional reduction ($r=0$). 
Thus at the one--loop level the ``$\epsilon$--scalar'' technique is 
equivalent to performing the algebra in the numerator of the integrand 
in four dimensions and making the integration in $D$ dimensions, 
or equivalently taking $D = 4 - r\epsilon$ with $r \to 0$, 
as we did in our calculations.

\section*{Acknowledgements}

We are very grateful to Prof. J. Sol\`a for the invitation to this 
interesting workshop. We also appreciated very much the smooth organization. 
In particular, we enjoyed the intimate and inspiring character of this 
workshop. This
work was supported by the ``Fonds zur F\"orderung der wissenschaftlichen
Forschung'' of Austria, project no. P10843--PHY.

\section*{References}

\end{document}